\documentclass[letterpaper,11pt]{article}
\usepackage{a4wide}
\usepackage[english]{babel}
\usepackage{authblk}
\usepackage{caption}
\usepackage{amsfonts}
\usepackage{amsmath}
\usepackage{amsthm}
\usepackage{amsbsy}
\usepackage[utf8]{inputenc} 
\usepackage{booktabs} 

\def\includegraphics{}
\usepackage{graphicx}
\usepackage{amssymb}

\title{Differential analysis of biological networks}
\author[1]{Da Ruan}
\author[1]{Alastair Young}
\author[1,2]{Giovanni Montana\footnote{Corresponding author: \tt{giovanni.montana@kcl.ac.uk}}}
\affil[1]{Department of Mathematics, Imperial College London, London SW7 2AZ, UK}
\affil[2]{Department of Biomedical Engineering, King's College London, St Thomas' Hospital, London SE1 7EH, UK}
\date{} 

\begin{document}

\maketitle

\abstract{In cancer research, the comparison of gene expression or DNA methylation networks inferred from healthy controls and patients can lead to the discovery of biological pathways associated to the disease. As a cancer progresses, its signalling and control networks are subject to some degree of localised re-wiring. Being able to detect disrupted interaction patterns induced by the presence or progression of the disease can lead to the discovery of novel molecular diagnostic and prognostic signatures. Currently there is a lack of scalable statistical procedures for two-network comparisons aimed at detecting localised topological differences.  We propose the dGHD algorithm, a methodology for detecting differential interaction patterns in two-network comparisons. The algorithm relies on a statistic, the Generalised Hamming Distance (GHD), for assessing the degree of topological difference between networks and evaluating its statistical significance. dGHD builds on a non-parametric permutation testing framework but achieves computationally efficiency through an asymptotic normal approximation. We show that the GHD is able to detect more subtle topological differences compared to a standard Hamming distance between networks. This results in the dGHD algorithm achieving high performance in simulation studies as measured by sensitivity and specificity. An application to the problem of detecting differential DNA co-methylation subnetworks associated to ovarian cancer demonstrates the potential benefits of the proposed methodology for discovering network-derived biomarkers associated with a trait of interest. }

\section{Introduction}
Current efforts at understanding diseases rely on the ability to identify differences between healthy and affected tissues. A number of high-throughput platforms are now commonly used to compare genome-wide molecular profiles collected from large cohorts of healthy and diseased subjects  in search for patterns that differentiate between them. For instance, in cancer research, gene expression and DNA methylation profiles from diseased tissues are compared to those extracted from normal controls in order to identify groups of genes whose expression or methylation levels are significantly different, and consequently associated to the trait of interest. From a statistical modelling standpoint, the primary interest of these studies lies in detecting statistically significant changes in average gene expression or methylation values in a two-sample comparison. A number of standard statistical tests, which are generally applied in a univariate fashion, have been proposed for this task and generate candidate sets of genes for further investigation \cite{Tusher01}. Statistical methods have also been developed to assess whether these candidate genes are over-represented in pre-defined biological pathways or subnetworks within protein interaction networks \cite{Nacu07}.  These developments are based upon the principle that, in order to understand the roles of genes in complex diseases, genes need to be studied in the context of the regulatory systems they are involved in \cite{Ideker02,Keller09,Nacu07}. 

An alternative way of analysing genome-wide expression and methylation levels observed in a random sample consists of studying their interaction patterns, which are often represented in the form of networks \cite{Dhaeseleer:2000tl, Dehmer08}.  Network edges quantify the similarity in transcription activity between two genes \cite{Zhang05} or in DNA methylation between two CpG islands \cite{Yang2015aa}, respectively. The notion of similarity is usually measured by linear correlation, partial correlation or mutual information coefficients estimated from the sample data \cite{Zhang05, Carter04}. The networks arising in the two-sample setting above can then be compared to assess whether there are statistically significant differences in network topology that can be associated to the disease. The detection of markedly distinct interaction patterns across conditions may be indicative of local disturbances within known biological pathways, and can be taken as candidate biomarkers. For instance, as a cancer progresses, it has been observed that its signalling and control networks are subjected to re-arrangments which are advantageous for the cancer \cite{Barabasi04}. Changes in methylation levels are believed to be among the earliest and most common alterations in human cancers \cite{Vaissiere:2009iy,Bartlett14}, and topological differences in healthy and diseased networks can reflect significant dysregulations associated to the disease \cite{Suzuki12}.

In this paper we discuss the the problem of comparing two labelled biological networks, each one representing a different population or condition, with the aim of detecting statistically significant differences between them. We approach this problem from a hypothesis testing perspective. This is a challenging statistical problem as only one random network is observed under each condition. Various computational methodologies have been developed to compare networks, including graph matching and graph similarity algorithms \cite{Brandes05}.  Graph matching algorithms have been used to discover similarities between molecular pathways across organisms and functions \cite{Di13,Yang07}, but are typically limited to unlabelled graphs, and are not concerned with hypothesis testing. Graph similarity algorithms also assume that the graphs are unlabelled, and  the attention has mostly focused on detecting patterns that are  most similar between networks \cite{Przulj07}. For instance, gene modules can be identified separately in each network first, and then compared across networks \cite{ Zhang05, Gill10, Zhu12}.  More closely related work includes inferential methods for performing two-sample hypothesis tests where the sampling unit is a network, and assess whether the two paired networks come from the same assumed model \cite{Yates2013}. 

We take a non-parametric approach to inference that does not require to make assumptions about a specific random network model. Our premise is that any true topological differences between the two networks would involve only a smaller set of edges, compared to all edges in the network, which we aim to detect. Our contributions to this problem are as follows. First, we consider the issue of choosing a distance measure between two paired networks that is able to capture subtle topological differences. Second, we discuss how to establish whether large values of this distance can be deemed statistically significant under a null hypothesis that the networks are independent. Finally, we ask whether it is possible to identify a differential subnetwork, starting from two large networks, in a computationally efficient manner.

The article is organised as follows. In Section \ref{GHD} we introduce a distance for labelled networks, the Generalised Hamming Distance (GHD). Building on this distance, a permutation-based test statistic for two-sample network comparisons is introduced in Section \ref{test}. Conditions for asymptotic normality are provided so that p-values can be obtained  in closed-form without the need to carry out computationally expensive permutations. In order to verify these results in special cases, in Section \ref{normality} we argue that the proposed conditions hold true for various random network models, and provide a sketch proof for the case of scale free networks. In Section \ref{subnetwork} we describe an algorithm, dGHD, for the detection of differential subnetworks. In Section \ref{sims} we present a number of simulation experiments that highlight the advantages of the proposed methodology under different graph models. As an illustrative application of the proposed methodology, a case-control study involving DNA co-methylation networks in ovarian cancer is presented in Section \ref{app}. We conclude with a discussion in Section \ref{discuss}.


\section{The generalized Hamming distance} \label{GHD}

We assume to have observed two paired biological networks, each represented by a graph, denoted by $\mathcal{A}=(V,E_A)$ and $\mathcal{B}=(V,E_B)$, respectively. Both graphs are defined on a common set, $V=\{1,2,\ldots,N\}$. The respective sets $E_A$ and $E_B$ of edges indicate the connection between the nodes in the two graphs. We also let the matrices $A = (A_{ij})$ and $B = (B_{ij})$ denote the two $(N \times N)$ adjacency matrices associated with graphs $\mathcal{A}$ and $\mathcal{B}$, respectively. 

The Hamming distance (HD) between $\mathcal{A}$ and $\mathcal{B}$ provides a commonly used metric to quantify the difference between the networks, and is defined by $\frac{1}{2} \text{tr}[(A-B)^2]$, where $\text{tr}[\cdot]$ denotes the trace of a matrix. This distance takes into account the number of edges that are in common between the two networks. Here we propose an extension of this metric, which we call the Generalised Hamming Distance (GHD), defined as
 \begin{equation}\label{ghd}
\text{GHD}(\mathcal{A},\mathcal{B}) = \frac {1}{N(N-1)} \sum_{i,j}  ( a'_{ij} - b'_{ij} )^2,
\end{equation}
where $a'_{ij}$ and $b'_{ij}$ are mean-centred edge weights defined as
$$
a'_{ij} = a_{ij} - \frac{1}{N(N-1)}\sum_{i,j}a_{ij}, \quad b'_{ij} = b_{ij} - \frac{1}{N(N-1)}\sum_{i,j}b_{ij}
$$
and $\sum_{i,j}$ denotes summation over distinct $i$ and $j$. The edge weights, which depend on the topology of the networks, provide a measure of connectivity between every pair of nodes $i$ and $j$ in $\mathcal{A}$ and $\mathcal{B}$, respectively.  When $a_{ij}$ and $b_{ij}$ are binary values indicating the presence or absence of an edge, i.e. are the elements of $A$ and $B$, $\text{GHD}(\mathcal{A},\mathcal{B})$ is related to the HD.  The specific node weights we propose here instead quantify the topological overlap (TO) between a pair of nodes by taking into account the local neighbourhood structure around those nodes \cite{Horvath11}. In the literature, the TO measure has been successfully applied for the detection of communities in biological networks, and there is empirical evidence that it carries biological meaning \cite{Zhang05,Allen12}. 

We use the one-step TO between nodes $i$ and $j$ indicating whether they share direct connections to other nodes. The weights are obtained from the adjacency matrix as follows:
\begin{equation} \label{aij}
a_{ij} =
\frac{\sum_{l\ne i,j}A_{il}A_{lj}+A_{ij}}{\min(\sum_{l\ne i}A_{il}-A_{ij},\sum_{l\ne j}A_{il}-A_{ij}) +1},
\end{equation}
when $i \ne j$, and otherwise $a_{ij}=1$, and analogously for $b_{ij}$. The GHD sums the squared differences $(a'_{ij}-b'_{ij})^2$ over all pairs of nodes in the network.  Note that the term $\sum_{l\ne i,j} A_{il}A_{lj}$  is a count of all vertexes $(i,l,j)$ containing node pair $(i,j)$. This term measures the connectivity information of each $(i,j)$ pair plus their common one-step neighbours. The denominator in \eqref{aij} can be written as $\min(d_i,d_j) +1 - A_{ij}$, where $d_i$ and $d_j$ represent the node degrees of $i$ and $j$, respectively. It is roughly equal to the smaller of $(d_i,d_j)$ and normalises $a_{ij}$ such that $0 \le a_{ij} \le 1$.  A large discrepancy between $a'_{ij}$ and $b'_{ij}$ indicates a topological difference localised around that pair of nodes.


\begin{figure}[!h]
\begin{center}
\includegraphics[scale=0.5, width=0.9\textwidth]{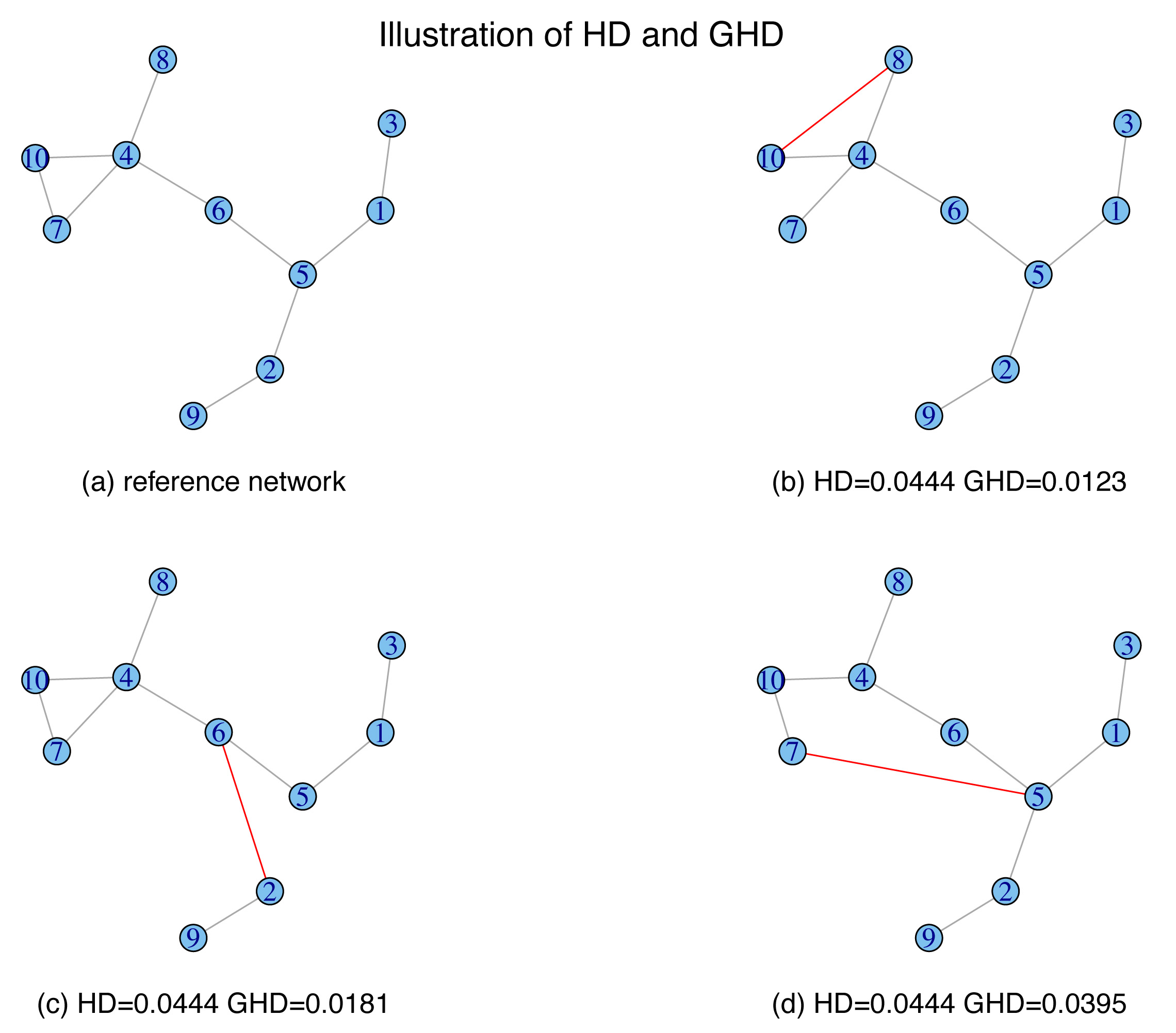}
\caption{Networks (b), (c) and (d) are generated from the reference network (a) by a single edge change. Both HD and GHD between the reference network and each modified paired network have been computed in each case.}
\label{fig:01}
\end{center}
\end{figure}


By exploring the neighbourhood of each node, the proposed GHD can detect subtle topological changes with higher sensitivity compared to the HD. A simple illustration of this is given in Figure \ref{fig:01}, where four simple networks are shown: the network labelled (a) is taken as reference while the three paired networks (b), (c), and (d) have been generated by changing the position of a single edge in (a). The two distances, HD and GHD, have been computed to quantify the difference between (a) and each of the other three networks. It can be observed that, whereas the HD is unable to distinguish between the three networks, the GHD score is more sensitive to subtle topological variations and can discriminate between them.

\section{A non-parametric test for network comparison} \label{test}

For inferential purposes, we require computing the probability that a distance as extreme or more extreme than the observed GHD value could have been observed by chance only. By treating the GHD as a random variable with unknown sampling distribution, this probability can be estimated non-parametrically via permutation testing. First, we specify the null hypothesis as being 
\begin{equation}
\mathcal{H}_0 : \text{networks $\mathcal{A}$ and $\mathcal{B}$ are independent.}
\end{equation}

By taking $\mathcal{B}$ as reference, each permutation consists of shuffling the labels of the nodes in $\mathcal{A}$ while keeping the edges unchanged. This generates a permuted network $\mathcal{A}_\pi$ that is isomorphic to $\mathcal{A}$, and the exchangeability property holds. In turn, this signifies that the original and permuted networks are generated from the same underlying, but unspecified, model \cite{Even72, Chang92}. Since all permutation networks are isomorphic, permuting  the labels of the network is equivalent to shuffling rows and columns of the adjacency matrix, an approach that bears some similarity with Mantel's test \cite{Mantel67} for the comparison of two distance matrices. All the the $N!$ possible permutations are then collected in a set $\Pi$, and for each $\pi \in \Pi$ a permuted GHD value is denoted as
$$
 \text{GHD}_\pi (\mathcal{A}_\pi,\mathcal{B}) =\frac {1}{N(N-1)} \sum_{i,j}(a'_{\pi{(i)}\pi (j)} - b'_{ij})^2, \nonumber
$$
and is calculated from the edge weights $a'_{\pi(i)\pi(j)}$ after permutation. 
The exact permutation distribution is obtained by carrying out an exhaustive calculation of all $\text{GHD}_\pi$ values, and p-values can then be evaluated as usual. In practice, however, doing so is computationally infeasible because the cardinality of $\Pi$ is generally extremely large, even for relatively small networks. The exhaustive evaluation for all permutations in $\Pi$ could be replaced by a Monte Carlo approach whereby only a smaller number of random permutations are explored. Nevertheless, the overall computational costs remain high for networks of the moderately large sizes observed in applications or when this procedure has to be repeated several times, for instance when searching for a differential subnetwork as in Section \ref{subnetwork}.

In what follows, we propose an alternative approach that removes the need to carry out computationally expensive permutation testing altogether. We demonstrate that, under our null hypothesis, the exact GHD permutation distribution can be approximated well by a normal distribution with moments that can be obtained analytically, in closed form. First, we notice that the GHD can be rewritten in an equivalent form in terms of a generalised correlation coefficient as follows:
\begin{eqnarray} \label{GHD2}
 \text{GHD}_\pi (\mathcal{A}_\pi,\mathcal{B})  = c  - \frac {2}{N(N-1)} \sum_{i,j} a'_{\pi(i)\pi(j)} b'_{ij},
\end{eqnarray}
where $c$ is a constant that does not change under permutations. By making use of this alternative representation, we are able to exploit well-known sufficient conditions for asymptotic normality, which can also be easily checked in practice. For a generalised correlation coefficient of this form, the exact permutation distribution is asymptotically normal under two sufficient conditions \cite{Daniels44,Friedman83,Pham89}:
\begin{subequations}
\begin{align}
\sum_{i,j} a'_{ij} & = \sum_{i,j} b'_{ij} =  0 \quad \text{ and } \label{conda} \\
\lim_{N\to\infty} & \frac{[\sum_{ijkl}a'_{ij}a'_{ik}a'_{il}]^2}{[\sum_{ijk}a'_{ij}a'_{ik}]^3}=\lim _{N\to\infty} \label{condb}  \frac{[\sum_{ijkl}b'_{ij}b'_{ik}b'_{il}]^2}{[\sum_{ijk}b'_{ij}b'_{ik}]^3}=0.
\end{align}
\end{subequations}

Condition \eqref{conda} follows directly from the definition of $a'_{ij}$ and $b'_{ij}$ as being mean-centred. In order to gain some insight into the meaning of condition \eqref{condb} in our context, it is instructive to consider the case where $a_{ij}$ and $b_{ij}$ are elements of the two adjacency matrices, i.e. they indicate the presence of an edge. On defining $a_{i\cdot}=\sum_{j \neq i} a_{ij}$ and  $\bar a = \frac{1}{N}\sum_{i}a_{i\cdot }$, we have 
\begin{equation} \label{nodeweight}
a'_{i\cdot} = \sum_{j \neq i} a'_{ij} = a_{i\cdot}-\bar a,
\end{equation}
and condition \eqref{condb}, with reference to network $\mathcal{A}$, can be written as
\begin{equation} \label{eq:rawcond2}
\lim _{N\to\infty} \frac{[\sum_i(Na'_{i\cdot})^3]^2}{[\sum_i(Na'_{i\cdot})^2]^3}=  \lim _{N\to\infty} \frac{[\sum_{i}(a_{i\cdot} - \bar a )^3]^2}{[\sum_{i}(a_{i\cdot} - \bar a )^2]^3}=0,
\end{equation}
and analogously for $\mathcal{B}$. It can be observed that, when using the adjacency matrix, $a_{i\cdot}$ represents the degree of the $i^{th}$ node. An analogous condition also applies to $\mathcal{B}$. Therefore, checking \eqref{condb} amounts to computing the degree of each node in the two networks, and assessing the limiting behaviour. When the TO measure is used instead, as in the GHD, the coefficient $a_{i\cdot}$ represents the overall topological overlap information at node $i$, and can also be computed using \eqref{nodeweight}.


When both \eqref{conda} and \eqref{condb} hold true, under the null hypothesis, the permutation distribution of $\text{GHD}(\mathcal{A},\mathcal{B})$ is approximately normal. We then standardise the GHD value by mean-centring and normalising it, so that it follows a standard normal distribution asymptotically,
\begin{equation} \label{GHD_test}
\frac{\text{GHD}_\pi(\mathcal{A}_\pi,\mathcal{B}) - \mu_{\pi}} {\sigma_{\pi} } \sim N(0,1)
\end{equation}
where $\mu_{\pi}$ and $\sigma_{\pi}$ are the mean and standard deviation of GHD under the exact permutation distribution, respectively. These two moments can be computed precisely and in closed-form by enumerative combinatorics; the calculations follow developments described in the context of related permutation-based testing procedures \cite{Mantel67}, and can also be found in \cite{Ruan2014}. Here we provide explicit formula for both $\mu_\pi$ and $\sigma^2_\pi$ as follows. First, we need to define   
\begin{eqnarray}
^tS_{a} = \sum_{i=1}^N\sum_{j=1}^N a^t_{ij} ~~ {\rm }~ t=1,2 &\text{and} & T_{a} =  \sum_{i=1}^N(\sum_{j=1}^N a_{ij})^2 ~~ \nonumber \\
^tS_{b} = \sum_{i=1}^N \sum_{j=1}^N b^t_{ij} ~~ {\rm}~ t=1,2 &\text{and}& T_{b} = \sum_{i=1}^N(\sum_{j=1}^N b_{ij})^2 ~~  \nonumber 
\end{eqnarray} where $a^t_{ij}$ and $b^t_{ij}$ are edge weights with power $t$. Here $\frac{^1S_a}{N(N-1)}$ and $\frac{^2S_a}{N(N-1)}$ are empirical raw moment of edge weight $a_{ij}$, and analogously for $b_{ij}$.
Furthermore we need to introduce the following quantities, 
\begin{eqnarray}
A_a = (^1S_a)^2,&& B_a = T_a - (^2S_a),~~\text{and}~~ C_a = A_a+2(^2S_a) -4T_a \nonumber \\
A_b = (^1S_b)^2,&& B_b = T_b - (^2S_b~),~~\text{and}~~ C_b = A_b+2(^2S_b) -4T_b \nonumber 
\end{eqnarray} Then, closed-form expressions for the mean $\mu_\pi$ and variance $\sigma^2_\pi$ are, 
\begin{eqnarray}
&&\mu_\pi = \frac{{^2S_a} + {^2S_b}}{N(N-1)} - \frac{2({^1S_a})({^1S_b})}{N^2(N-1)^2} \nonumber\\
&&\sigma^2_\pi = \frac{4}{N^3(N-1)^3}[2({^2S_a} )( {^2S_b})+\frac{4(B_a)(B_b)}{N-2} + \frac{(C_a)(C_b)}{(N-2)(N-3)} - \frac{(A_a)(A_b)}{N(N-1)}] \nonumber
\end{eqnarray} With the expressions for the first two exact moments, a corresponding p-value can therefore be efficiently computed from the normal approximation, even for very large networks. We will exploit the computational efficiency gained here in Section \ref{subnetwork}, where we apply the test repeatedly on networks of increasingly smaller size in order to detect differential subnetworks.

\section{Validation of asymptotic normality on scale-free networks} \label{normality}

The closed-form approximation for the computation of p-values only requires that conditions \eqref{conda} and \eqref{condb} are satisfied, and does not need any random network model to be specified. These two conditions can also be verified analytically in special case when certain random network models are assumed. For instance, in \cite{Ruan2014} it was proved that these conditions hold true for scale-free (SF), random geometric (RG) and Erd\"os-R\'enyi  (ER) network models when using both HD and GHD distances. In this section we provide a simplified  proof for the case of SF networks using the Hamming distance. This proof should serve as an illustration of how these derivations can be carried out analytically, and as simple validation of the methodology described in Section \ref{test} for SF  networks. An analogous proof using the GHD distance can be found in the Supplementary Material, and we refer the reader to  \cite{Ruan2014} for the other models.

A SF network is a network whose node degree distribution follows a power law, at least asymptotically, and has often been used to describe real biological networks \cite{Chung03,Van04,Jordan04}. The degree of each node is assumed to be an independent and identically distributed (IID) random variable with probability mass function defined as
\begin{equation} \label{eq:powerlaw}
P(d_i = k) =  ck^{-\alpha}, \quad k = m,m+1...,K,
\end{equation}
where $m$ and $K$ are the lower and upper cut-offs for the node degree, respectively, $c$ is a normalising constant, and $\alpha$ represents a power exponent. It is generally assumed that $\alpha$ is greater than 1, and the lower cut-off $m$ is generally be taken to be $1$. The upper cut-off $K$ for $\alpha > 2$ is conventionally specified as $K = N^{\frac{1}{\alpha-1}}$
\cite{Cohen01}, and generally $K= N-1$ for $1 <\alpha \le 2$.  Values of $\alpha$ for different biological networks have been characterised, and mostly vary between $1.4$ to $1.7$ \cite{Chung03}.

On defining the weights $a_{ij}$ and $b_{ij}$ as elements of $A$ and $B$, respectively, \eqref{eq:rawcond2} becomes
\begin{eqnarray} \label{eq:adjcond}
 \lim _{N\to\infty} \frac{[\sum_{i}(d_{i} - \bar d )^3]^2}{[\sum_{i}(d_{i} - \bar d )^2]^3}=0,
\end{eqnarray} where $\bar d$ is the average node degree. In order to study this limiting behaviour, we exploit the fact that both numerator and denominator are powers of the centralised empirical moments of the node degree distribution. We let $\mu_s =  c \sum_{d=1}^{K} d^{s-\alpha}$ denote the $s^{th}$ theoretical moment and $m_s=\frac {1}{N}\sum_{i=1}^Nd_i^s$ the corresponding empirical moment of this distribution. In order to study the limit above we need to characterise the order of $m_s$, for $s=1,2,3$, as $N$ increases. Our strategy here consists of first characterising the order of $\mu_s$ asymptotically, for the first three moments, and establishing a correspondence with $m_s$.

We start by examining the order of $\mu_s$, for $s=1,2,3$, in the limit. Since this depends on $s$, we consider three distinct cases: (a) $s-\alpha +1 < 0$, (b) $s-\alpha +1 = 0$ and (c) $s-\alpha +1 > 0$. For (a), the order of $\mu_s$ is
$\sum_{d=1}^{K} \frac{1}{\alpha - 1} d^{-1} = O(1)$. For (b), the order of $\mu_s$ is $\sum_{d=1}^K d^{-1} = O(\ln(K))$. Finally, for (c), we need to study how $\mu_s$ increases with $K$. First, we apply the Euler-Maclaurin formula,
$$
 \sum_{d=1}^K d^{s-\alpha} = K^{s-\alpha+1} + (\alpha-s)\int_{1}^{K} \frac{\left\lfloor x \right\rfloor}{x^{\alpha-s+1}} dx + O(1),
$$
where $\lfloor x \rfloor$ denotes the largest integer that is not greater than $x$. To compute the order of $\sum_{d=1}^K d^{s-\alpha} $, we need to know which one of the two terms in the sum dominates in order. By applying l'Hospital's rule we have
\begin{eqnarray*}
\lim_{K\to\infty}\frac{s\int_{1}^{K} \frac{\left\lfloor x \right\rfloor}{x^{\alpha -s+1}} dx}{ K^{s-\alpha+1}}
&=& \frac{s}{s-\alpha+1},
\end{eqnarray*}
which is a finite constant, and hence $\mu_s$ has the same order as $K^{s-\alpha+1}$. For a SF network,  the condition for asymptotic normality also depends on the values taken by the exponent. In the case where $1 < \alpha <2$, for which $K=N-1$, the calculation of the $s^{th}$ moment falls under case ($c$), hence we conclude that the order of the first three theoretical moments are, respectively, $O(N^{2-\alpha}), O(N^{3-\alpha})$ and $O(N^{4-\alpha})$.

We now turn to the direct comparison of the orders of $\mu_s$ and $m_s$ in the limit. Specifically, we assess whether the order of each $\mu_s$ established above also holds true for the corresponding $m_s$. This can be verified by checking that
\begin{eqnarray}\label{sameorder}
\lim_{N \to \infty} \frac{m_s}{\mu_s} = c_s,  
\end{eqnarray}
for $s=1,2,3$, and for some positive constants $c_s$. To study the above limit, we apply the Weak Law of Large Numbers (WLLN). For the WLLN to hold, $\mu_s$ must be finite. Hence we first transform $d_i$ so that $\mu_s$, after the transformation, is finite. We let $N_s = N^{\frac {s+1-\alpha} {s}}$, and define $z_{si} = \frac {d_i}{N_s}$ . The distribution of $z_{si}$ is 
 \begin{equation*}
P(z_{si}=z) = c'z^{-\alpha},  \qquad z = \frac {1}{N_s},\frac {2}{N_s} ,..,\frac {K}{N_s},
\end{equation*}
where $c' = c{N_s}$. Thus the $s^{th}$ theoretical moment of $z_{si}$ is
\begin{eqnarray*}
\mu_{zs} = c'\sum_{z} z^{s-\alpha}= c' \sum_{d}[\frac{d}{N_s}]^s d^{-\alpha}= \frac{\mu_s}{N^{s+1-\alpha}},
\end{eqnarray*} which is finite.
Denoting by $m_{zs}$ the $s^{th}$ empirical moment of $z_{si}, i = 1,...,N$, we have
\begin{eqnarray*}
m_{zs} = \frac{1}{N} \sum_{i=1}^N z_{si}^s = \frac{1}{N} \sum_{i=1}^N (\frac{d_i}{{N_s}})^s = \frac{m_s}{N^{s+1-\alpha}}.
\end{eqnarray*}
Now, since $\mu_{zs}$ is finite and since $d_1, d_2,...,d_N$ are assumed IID, $z_{s1}, z_{s2},...z_{sN}$ are also IID, and according to the WLLN, $m_{zs} $ converges to $\mu_{zs} $ in probability. Hence we have
\begin{eqnarray*}
1 = \lim_{N \to \infty} \frac{m_{zs}}{\mu_{zs}} = \lim_{N \to \infty} \frac{\frac{m_s}{N^{s+1-\alpha}} }{\frac{\mu_s}{N^{s+1-\alpha}}} = \lim_{N \to \infty} \frac{m_s}{\mu_s},
\end{eqnarray*}
indicating that $m_s$ and $\mu_s$ are of the same order asymptotically. Using this result, we are able to approximate the orders of the numerator and denominator of condition (\ref{eq:rawcond2}):
$
\sum_{i}(d_{i} - \bar d )^3 = N(m_3-2m_2m_1+2m_1^3)
$
is $O(N^{4-\alpha+1})$, and $
\sum_{i}(d_{i} - \bar d )^2 = N(m_2 - m_1^2)
$ is $O(N^{3-\alpha+1})$. Substituting into (\ref{eq:rawcond2}), we see that the numerator is of order $O(N^{8-2\alpha+2})$, the denominator is of order $O(N^{9-3\alpha+3})$, and therefore the ratio is of order $O(N^{\alpha -2})$. Hence for $1 < \alpha < 2 $, the limit in (\ref{eq:adjcond}) is $0$. By following a similar procedure, it can be proved that the normality condition is also satisfied when $\alpha \ge 3$. 

\section{Differential subnetwork detection} \label{subnetwork}


In this section we leverage the test statistic of Section \ref{test} to detect a differential subnetwork. When comparing the two networks, the expectation is that only a subset of edges would present altered interaction patterns. This task is formulated here as the problem of detecting a subset $V^* \subset V$ for which there is no sufficient evidence to reject the null hypothesis that the corresponding subnetworks $\mathcal{A}^*(V^*,E_{A^*})$ and $\mathcal{B}^*(V^*,E_{B^*})$ are statistically independent.

An algorithm for the detection of $V^*$ should take into account the fact that a certain degree of topological difference between $\mathcal{A}$ and $\mathcal{B}$ is always bound to be observed, even when the two population networks are the same, due to finite sample variability. The GHD test provides an efficient way to assess the statistical significance of any observed discrepancy between two paired networks, and is used as a building block to derive an algorithm that identifies differential subnetworks.

We indicate by $V_K$ a subset of $V$ of size $K \leq N$, and define the centralised GHD test statistic computed by comparing  $\mathcal{A}=(V_K,E_A)$ and $\mathcal{B}=(V_K,E_B)$ by
\begin{equation} \label{centralGHD}
\Delta_{V_K} = \text{GHD}(\mathcal{A}(V_K,E_A),\mathcal{B}(V_K,E_B)) - \mu_{V_K,}
\end{equation}
where $\mu_{V_K}$ is the mean of the permutation distribution for node set $V_K$.
Furthermore we define $\Delta_{V_K|i}$ to be the centralised GHD value computed by comparing the two networks after removal of node $i$. The quantity
$$
\delta_i = \Delta_{V_{K}|i} - \Delta_{V_K},
$$
measures the influence that node $i$ has on the mean-centred GHD test when comparing two subnetworks defined on set $V_K$. We propose an iterative procedure which removes a node or set of nodes at each step, and generates a sequence of node sets of increasing smaller size, i.e.
$$
V_N \supset V_{N-1} \supset \ldots \supset V_{N_{\text{min}}},
$$
where $N_{\text{min}} < N $ is a constant indicating the smallest allowed size of subnetwork. Starting with $V_N$, the two corresponding networks are compared by the GHD test, and a p-value is computed, as described previously. For each node indexed by $i = 1,...,N $, the corresponding $\delta_i$ is computed, and the node associated with the largest positive $\delta_i$ value is removed. Given a new set $V_{N-1}$, the process is then repeated again, and then again until a specified minimal set size is reached.

\begin{figure}[!h]
\begin{center}
\includegraphics[scale=0.5, width=0.9\textwidth]{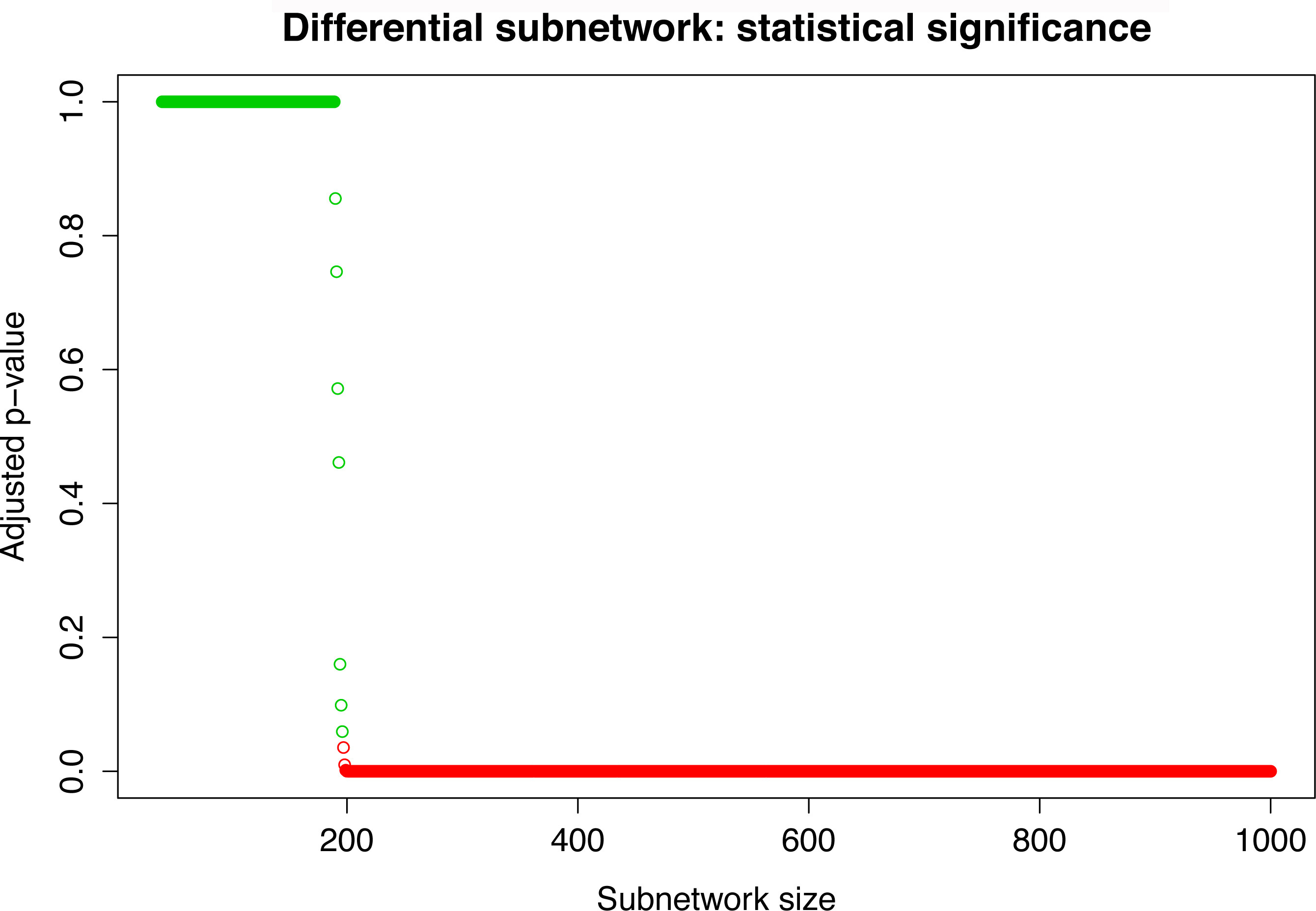}
\caption{Sequence of adjusted p-values produced by the dGHD algorithm as a function of subnetwork size. The size of the subnetwork is progressively reduced by removing nodes that further increase the distance between the subnetworks. For this example we simulated 2D RG networks of size 1,000 and subnetworks of size 200.}\label{subnet_power}
 \end{center}
\end{figure}


This simple algorithm produces a monotonic sequence of $p$-values that increases as the subnetwork size decreases (e.g. see Figure \ref{subnet_power}). The p-values should be adjusted for multiple testing, e.g. by controlling the false discovery rate \cite{Benjamini01}. In the presence of a differential subnetwork, the sequence is expected to feature a peak corresponding to the size of the subnetwork. Specifically, for a given desired significance level $\alpha$, the algorithm finds the largest $K$, with $N \geq K \geq N_{\text{min}}$, such that the adjusted p-value exceeds $\alpha$. Clearly the algorithm benefits from the fact that p-values at each iteration can be computed very quickly in closed-form. 

\section{Simulation experiments} \label{sims}

In this section we report on three different simulation experiments that have been carried out to study the properties of the proposed methodology. Our simulations make use of RG networks, which are plausible models for biological networks \cite{Przulj04,Przulj07,Ay07,Horvath08}. Two-dimensional RG networks were generated by first uniformly sampling $N$ points on $[0,1]^2$, each one corresponding to a node in the graph. A pair of nodes was connected by an edge if the Euclidean distance between the corresponding two-dimensional points was smaller than a pre-determined threshold $d$.  

\begin{figure}[!h]
\begin{center}
\includegraphics[scale=0.5, width=0.8\textwidth]{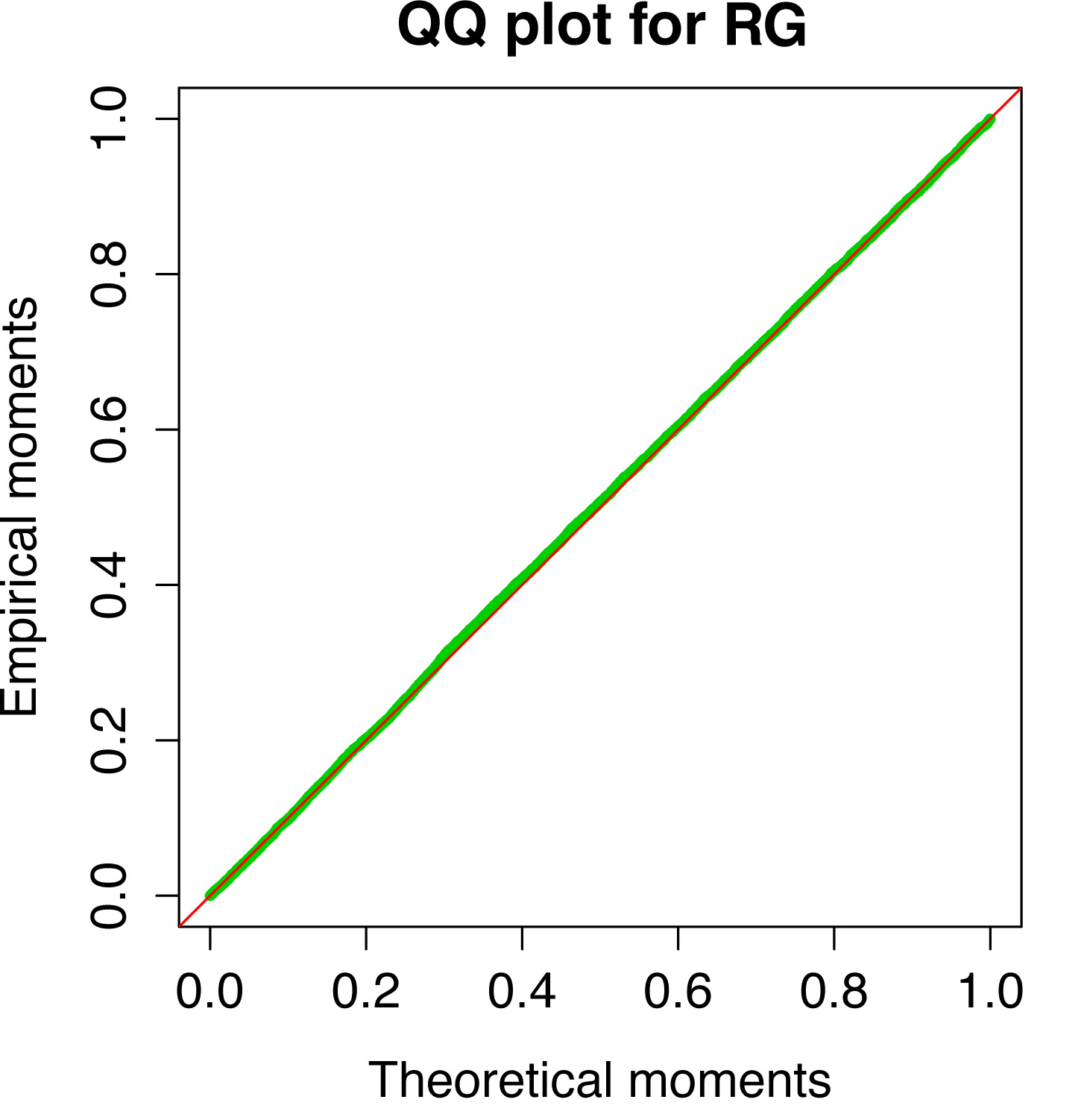}.
\caption{\footnotesize  QQ plots confirming the asymptotic null distribution of the GHD test.  For RG model, 10,000 paired networks of size $250$  were independently generated and the GHD test was applied to detect differences between them.  An empirical distribution of p-values was obtained through $10,000$ comparison for each model, which under the null is expected to be uniformly distributed on $[0,1]$. The figure shows that empirical and theoretical moments agree. } \label{fig:QQGHD} 
\end{center}
\end{figure}

The purpose of the first simulation study was to confirm the asymptotic null sampling distribution of the GHD statistic. In this case we randomly generated $10,000$ pairs of networks $\mathcal{A}$ and $\mathcal{B}$ of size $N=250$, with parameters  $d=0.3$ and $d=0.15$. For each $d$ value, paired networks were independently generated, and the GHD test was computed to detect differences between them. As a result of this process, we obtained an empirical distribution of p-values. Under the null, this distribution is expected to be uniform on $[0,1]$, and the resulting QQ plots confirm that the empirical moments of this distribution agree perfectly with the expected theoretical moments for a RG model; see Figure \ref{fig:QQGHD}.

In the second study, we compared the ability of the GHD test to detect differential networks against three competing tests: Mean Absolute Difference (MAD) \cite{Butts01}, Quadratic Assignment Procedure (QAP) \cite{Hubert87} and Conditional Uniform Graph (CUG) \cite{Anderson99}.  The MAD test counts the number of different edges in the two networks 
\begin{equation} \label{mad_test}
\text{MAD}(\mathcal{A},\mathcal{B}) =  \frac{1}{N(N-1)}\sum_{i,j}|a_{ij} - b_{ij}|,
\end{equation}
where $a_{ij}$ and $b_{ij}$ correspond to the $(i,j)$ elements in the adjacency matrices of $\mathcal{A}$ and $\mathcal{B}$, respectively. 
The QAP uses edge set product statistics to test for the independence between networks,
\begin{equation} \label{QAD}
\text{QAP}(\mathcal{A,B}) =  \frac{1}{N(N-1)} \sum_{i,j} a_{ij} b_{ij} ,
\end{equation}
where $a_{ij}$ and $b_{ij}$ are again elements of the adjacency matrices. For both the MAD and QAP tests we also used the traditional permutation testing approach. We further included in the study the CUG approach. According to this procedure, random networks are generated with pre-determined properties, such as size and density, matching the properties of the observed networks. For each simulated pair of random networks, a measure of correlation between networks is computed, and its empirical distribution is built up over many simulations. The correlation coefficient is defined as:
\begin{eqnarray*}
\text{gcor}(\mathcal{A},\mathcal{B}) = \sum_{i,j}(a_{ij}-\frac{\sum_{i,j}a_{ij}}{N(N-1)})(b_{ij}-\frac{\sum_{i,j}b_{ij}}{N(N-1)}),
\end{eqnarray*}
where $a_{ij}$ and $b_{ij}$ are elements of the adjacency matrices for $\mathcal{A}$ and $\mathcal{B}$, respectively \cite{vanWijk10}.

\begin{figure}[!h]
     \includegraphics[scale=0.5, width=0.9\textwidth]{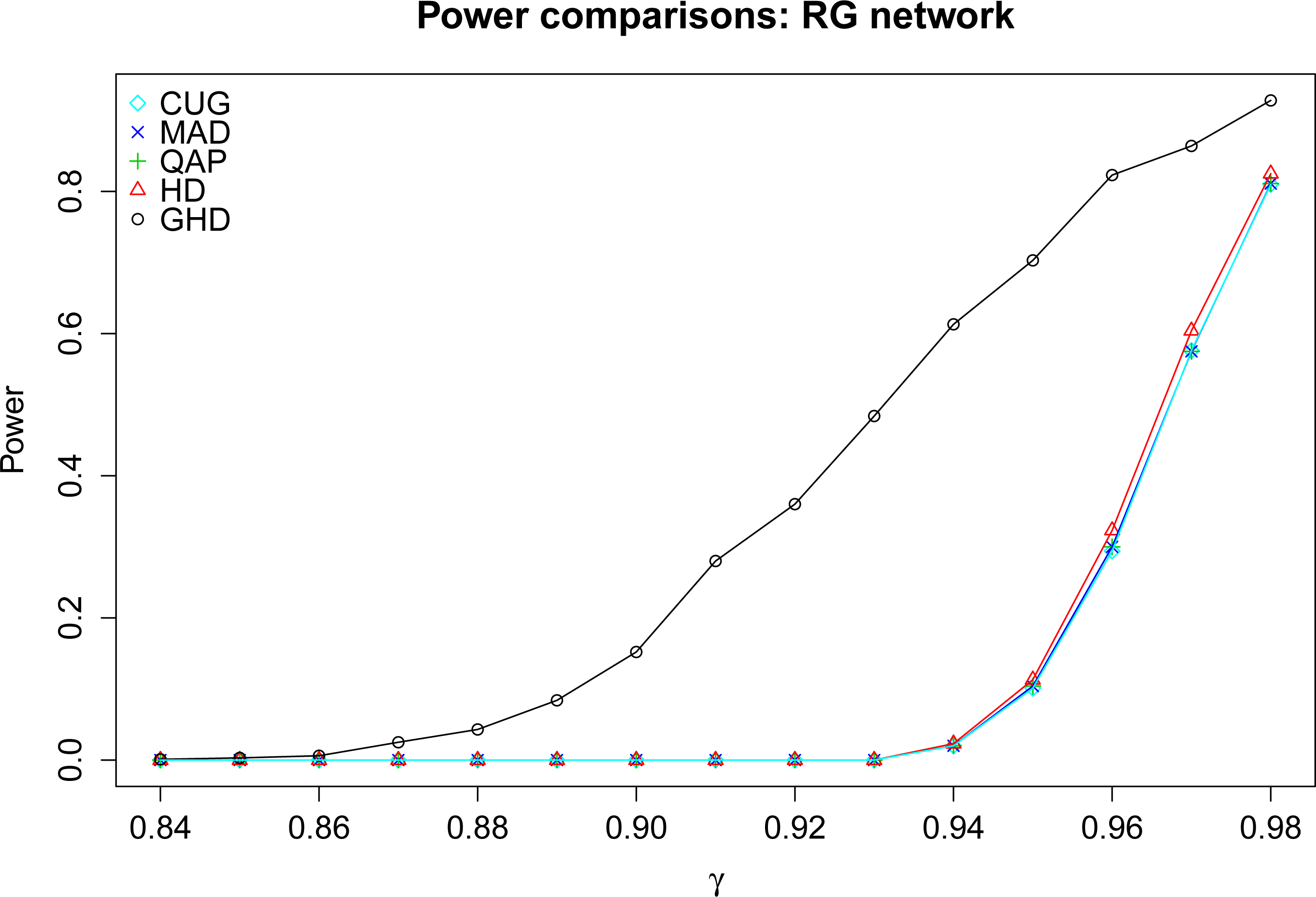}
\caption{\footnotesize The "power" is defined as the proportion of replications when the null hypothesis of independence is not rejected. As the noise level $\gamma$ increases, the GHD test has more power to detect true structural changes compared to competing methods. Simulations are based on 2D RG networks.} \label{fig:noise_example1}
\end{figure}

This experiment required the simulation of paired networks with a pre-specified degree of topological dissimilarity. This was achieved by generating $\mathcal{A}$ first, using one of the two random models as described above. Network $\mathcal{B}$ was then obtained by first making an exact copy of $\mathcal{A}$, and then randomly shuffling a fixed proportion $\gamma$ of edges so that, as $\gamma$ increases, the dissimilarity between $\mathcal{A} $ and $\mathcal{B}$ increases. For each given value of $\gamma$, we generated $1,000$ pairs of networks, computed the tests and corresponding p-values, and evaluated the proportion of tests that rejected the null hypothesis of independence at a $5\%$ significance level. The results of this study are summarised in Figure \ref{fig:noise_example1} where the "power" is defined as the the proportion of replications, out of $1,000$, when we accept the null hypothesis of independence. This rises from zero at $\gamma=0.84$, when networks are still associated, to close to $1$ when a lot of shuffling has been carried out, to produce nearly independent networks. This figure shows that for noise levels as large as $\gamma = 0.93$, the tests based on HD consider the two networks to be strongly associated. It is only when reaching that threshold that their power starts increasing rapidly away from zero. This suggests that the tests based on HD may be too stringent for real application and miss importance differential patterns. By contract, the GHD test is able to detect differences at lower noise levels compared to other tests and capture more subtle differences. This is not surprising as GHD is more sensitive to topological changes, as seen in Figure \ref{fig:01}.

In the third simulation study, we carried out an investigation to assess the behaviour of the differential subnetwork detection algorithm, and quantify its performance in comparison with other tests. We report on  experiments involving RG networks $\mathcal{A}$ and $\mathcal{B}$ of size 1,000 and generated as described above using a noise parameter $\gamma$.
Two independent subnetworks, denoted here by ${\mathcal{A}}^*$ and ${\mathcal{B}}^*$, were introduced by randomly selecting a subset $V^* \subset V$ of size $200$, and replacing the existing edges with connections simulated from two independent RG networks. For each value of $\gamma$, we generated $100$ such paired large networks containing smaller differential subnetworks.  We term a true positive (TP) a node that is correctly identified as belonging to the differential subnetwork, and a false negative (FN) a node that belongs to the subnetwork but has not been detected by the algorithm. Similarly we define false positives (FP) and true negatives (TN). In Table \ref{tab:fscore} we report the sensitivity or true positive rate (TPR) computed as TP/(TP+FN), and the specificity (SPC) computed as TN/(FP+TN). For comparative purposes, we have also implemented an alternative algorithm, called dHD, which is similar to dGHD but uses the Hamming distance instead for distance calculations. As can be observed, both dHD and dGHD maintain high sensitivity and specificity up to moderately high noise levels. For noise levels at the top end of the spectrum, dHD has slightly higher sensitivity but much smaller specificity than dGHD, indicating that it detects a larger number of incorrect nodes. 

\begin{figure}[!h]
\begin{center}
\includegraphics[scale=0.5, width=0.9\textwidth]{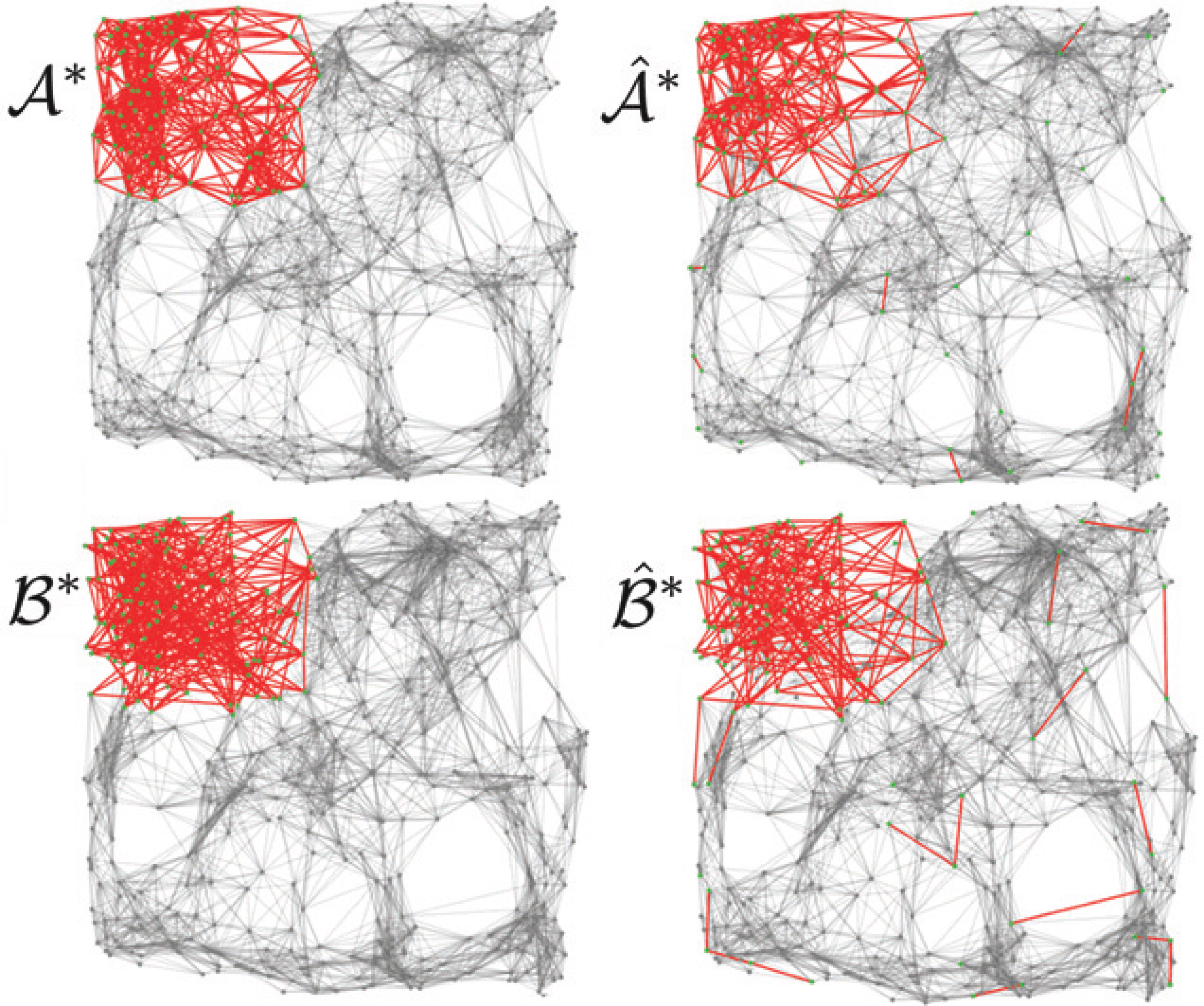}
\caption{\footnotesize {\bf } Example of differential subnetworks detected by dGHD using 2D RG networks. ${\mathcal{A}}^*$ and ${\mathcal{B}}^*$ are the true simulated independent subnetworks, and $\hat{\mathcal{A}}^*$ and $\hat{\mathcal{B}}^*$ are the subnetworks detected by the algorithm ($\gamma \approx  0.23 $). Nodes belongs to differential subnetwork are coloured in green, and edges colored red. Please refer to Table \ref{tab:fscore} for full results.} \label{fig:simnet}
\end{center}
\end{figure}

Figure \ref{fig:simnet} provides an example of simulated networks $\mathcal{A}$ and $\mathcal{B}$ and ground truth differential subnetworks  ${\mathcal{A}}^*$ and ${\mathcal{B}}^*$ as well as the differential subnetworks $\hat{\mathcal{A}}^*$ and $\hat{\mathcal{B}}^*$ detected  by dGHD in one of the 100 simulations. The corresponding sequence of p-values generated by running the dGHD algorithm in this example is shown in Figure \ref{subnet_power}. It can be noticed how the null hypothesis of independence is rejected for all the subnetworks of size ranging from 1000 down to 200, at which point there is no evidence to reject the null, and the algorithm produced large p-values for all sizes smaller than 200. 

\begin{table}[!h]
\centering
\caption{\textbf{Sensitivity (TPR) and specificity (SPC) of the subnetwork detection algorithms for different values of $\gamma$, the noise level. The results are based on simulated RG networks.}}\label{tab:fscore}
 \begin{tabular}{cccccccc}\hline
 & $\gamma$ & 0.055& 0.11 & 0.23 & 0.54   & 0.79 & 0.95  \\\hline
 dGHD  & TPR  & 0.897 & 0.889 & 0.855 & 0.627& 0.570 & 0.789  \\
 & SPC & 0.987 & 0.984 & 0.974 & 0.912& 0.768 & 0.439 \\
    \hline
  dHD & TPR &0.914 & 0.904 & 0.872 & 0.725& 0.712 & 0.862 \\
  & SPC & 0.978 & 0.971 & 0.956 & 0.843& 0.567 & 0.201 \\
\hline
 \end{tabular}
 \end{table}


\section{Application to co-methylation networks in ovarian cancer} \label{app}

We present an application to a case-control epigenetic study of ovarian cancer. The dataset for this study was originally presented in \cite{Teschendorff10}. Methylation profiles for $27,578$ CpGs islands were obtained from whole blood samples in $540$ women, of which $266$ were samples taken from postmenopausal women with ovarian cancer and $274$ were from age-matched healthy controls. In our analysis we set out to compare control and case DNA co-methylation networks in search of a differential subnetwork. 

\begin{figure}[!h]
\begin{center}
\includegraphics[scale=0.5, width=0.9\textwidth]{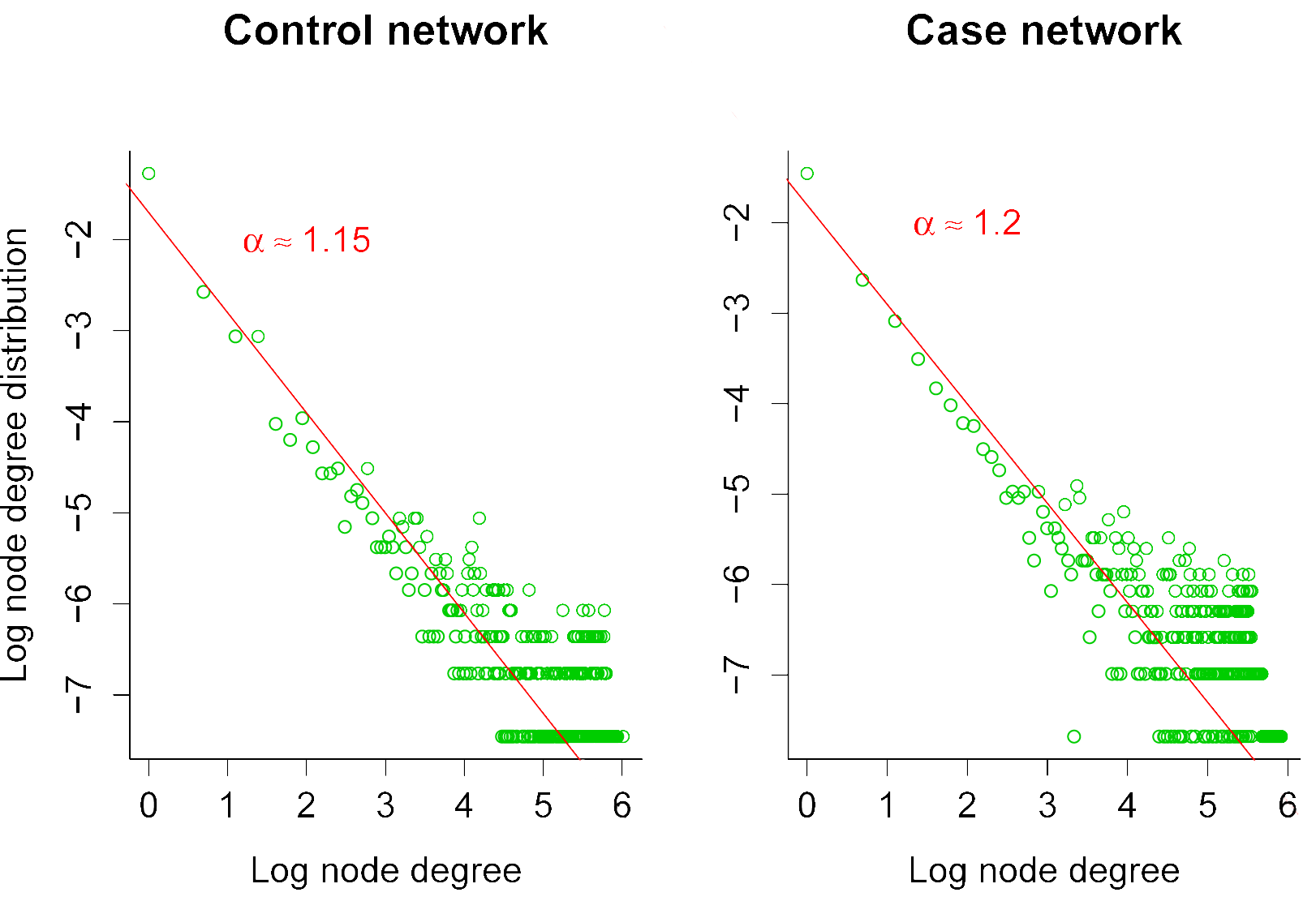}
\caption{\footnotesize  Node degree distribution for control and case co-methylation networks. Both plots shows that the networks roughly follow SF network models. } \label{fig:powerlaw}
\end{center}
\end{figure}

Raw data files were downloaded from GEO (repos. number GSE19711), and were obtained from Illumina Infinium 27k Human DNA methylation Beadchip v1.2. The raw data was pre-processed by using the {\tt lumi} package in R \cite{Du08}. After quantile normalization, PCA applied to the beta value was used to detect and remove extreme outliers. After quality control, $243$ control samples and $215$ case samples remained for further analysis. The networks was inferred by taking each probe as a node. Following \cite{Horvath12}, an adjacency measure was computed as $\omega_{ij}= | (1+\text{cor}(g_i,g_j))/2 |^b  $ where $\text{cor}(g_i,g_j)$ denotes the Pearson's correlation coefficient between beta values observed at the $i^{\text{th}}$ and $j^{\text{th}}$ CpG sites. The power exponent $b$ was set to a default value of 12 so as to place more emphasis on higher positive correlations \cite{Zhang05}. Two nodes were linked in the network if $\omega_{ij}$ was higher than $0.2$ so that the presence of an edge indicates a strong correlation. This value also yields networks that roughly follow a SF model (see Figure \ref{fig:powerlaw}). The number of resulting edges is $48,224$ and  $75,913$ in the control network $\mathcal{A}$ and case network $\mathcal{B}$, respectively. 

 \begin{figure}[!h]
      \includegraphics[width=0.9\textwidth]{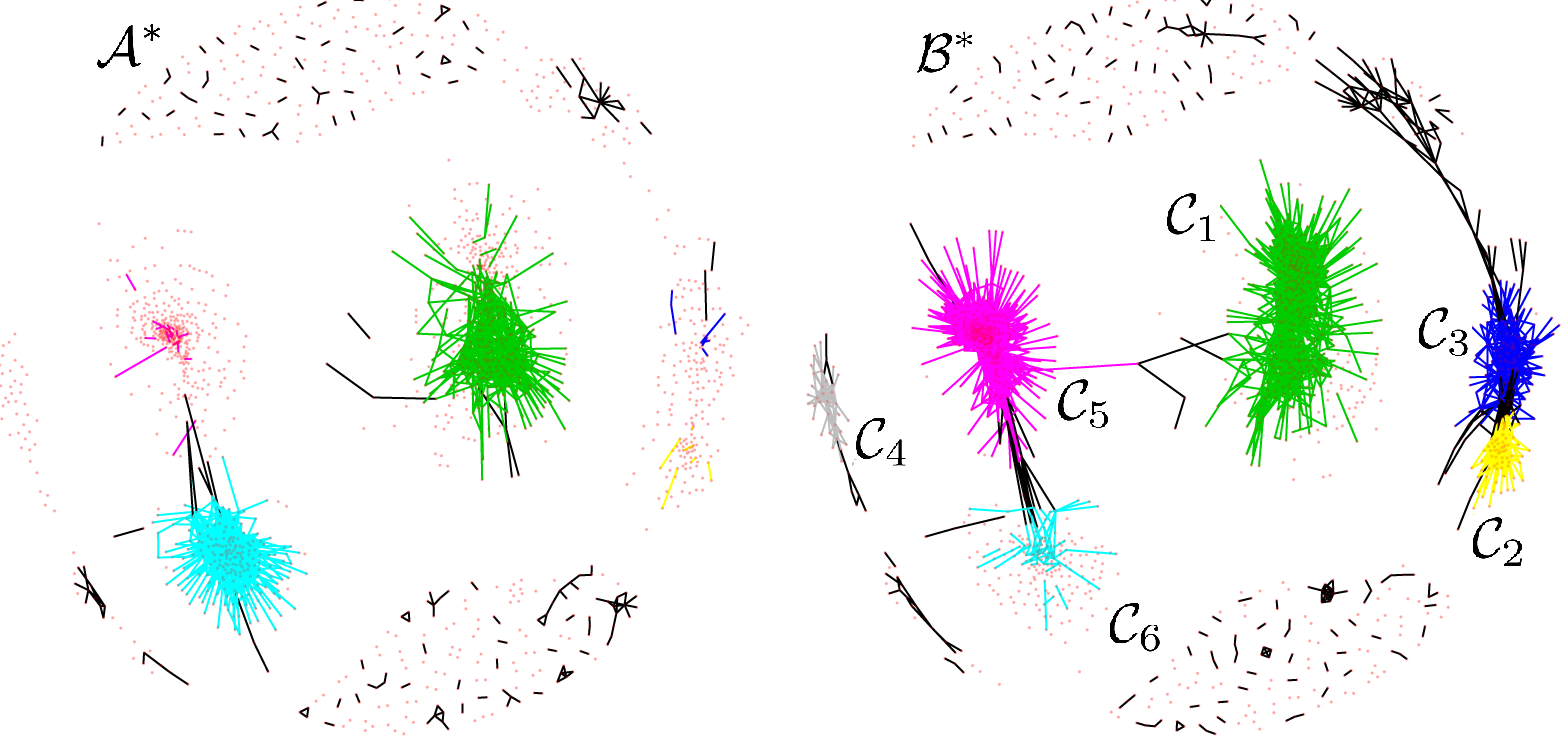}
\caption{\footnotesize DNA co-methylation networks: differential subnetworks $\mathcal{A}^*$ (controls) and $\mathcal{B}^*$ (cases) detected by dGHD algorithm. Six main communities within the subnetworks are characterised by a much higher network density in cancer patients compared to healthy controls. Differential methylation is mostly concentrated in $\mathcal{C}_2$, $\mathcal{C}_3$, $\mathcal{C}_5$ and $\mathcal{C}_6$ (See also Table \ref{community} and Figure \ref{fig:diffmethy}).} \label{fig:subnetworks}
\end{figure}


At a significance level of $5\%$ and after correction for multiple testing, the dGHD algorithm detected a subnetwork of size $1,642$, with $1,954$ edges in $\mathcal{A}^*$ and $12,556$ edges in $\mathcal{B}^*$. The two resulting subnetworks are presented in Figure \ref{fig:subnetworks}. Although the algorithm does not constrain the differential networks to be connected, they both comprise a number of connected subgraphs. The Walktrap community detection algorithm, as implemented in the R package {\tt iGraph} \cite{Pons06}, was used to identify communities in these two subnetworks, as shown in the Figure. The density of the six largest communities, which are denoted $\mathcal{C}_1, \ldots, \mathcal{C}_6$, differs quite substantially between control and cancer networks. In almost all communities, the density is much higher in $\mathcal{B}^*$, with the exception of $\mathcal{C}_6$, where it is higher in $\mathcal{A}^*$. 

To gain initial insight into the biological meaning of the subnetworks and the communities within them we used the R package {\tt GOstat} \cite{Beissbarth04} to identify enriched Gene Ontology (GO) terms within two broad categories, Biological Processes (BP) and Molecular Functions (MF). At a $5\%$ significance level, the hypergeometric test detected 762 BP and 154 MF statistically significant terms enriched in the subnetworks where most of these terms can be found in 6 communities. For instance, the top three BPs were response to stimulus, cellular response to stimulus and response to chemical stimulus, and the top three MFs were protein binding, collagen binding and RNA polymerase II transcription cofactor activity. Furthermore, we carried out a pathway enrichment analysis to identify any significantly enriched KEGG pathways. At a $5\%$ significance level, 12 pathways were found to be enriched, including hematopoietic cell lineage, acute myeloid leukemia, and regulation of action cytoskeleton. 

\begin{figure}
\includegraphics[width=0.9\textwidth]{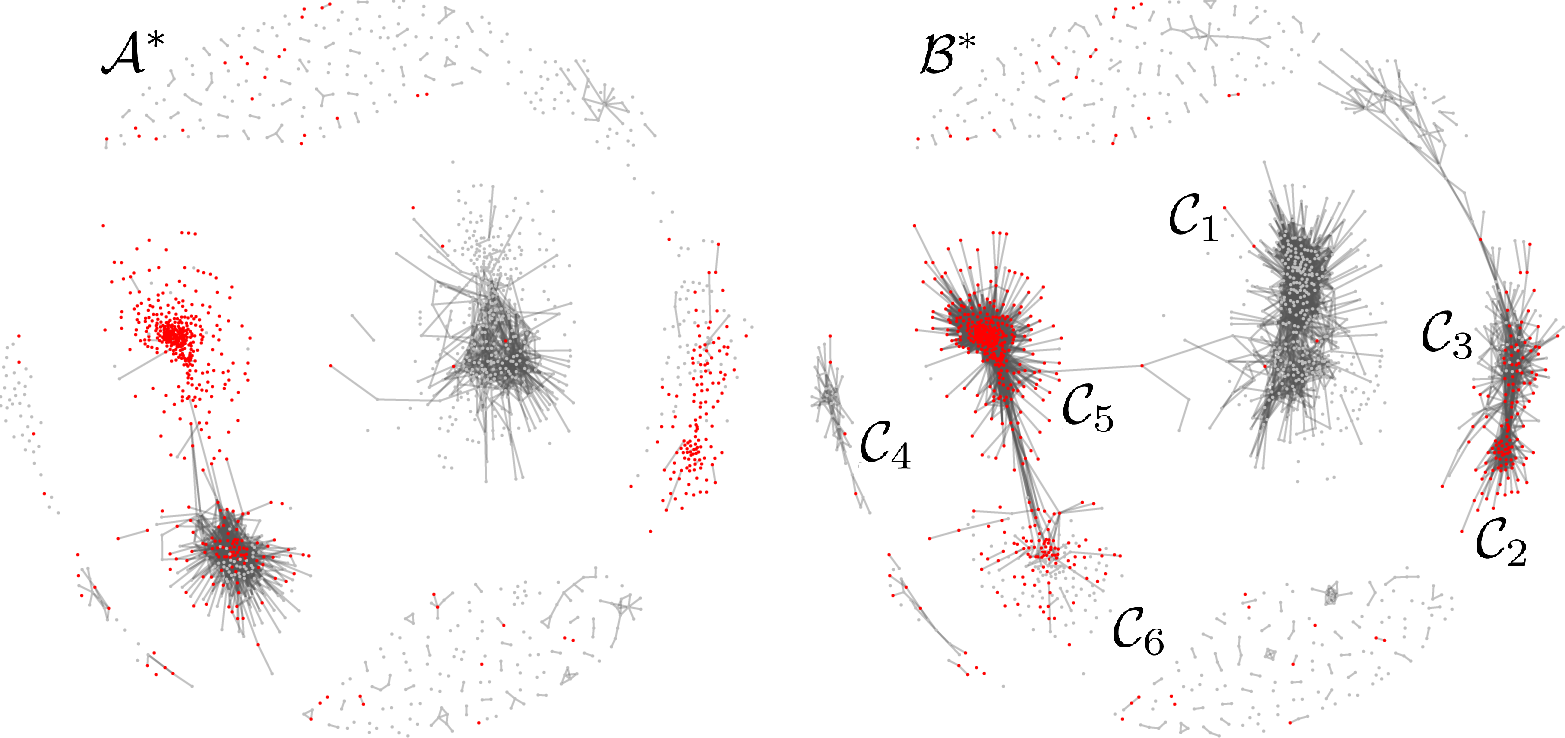}
\caption{\footnotesize Visualization of the distribution of differential methylated probes (red) in differential subnetworks detected by dGHD in the DNA co-methylation networks. } \label{fig:diffmethy}
\end{figure}

Probes showing statistically significant changes in mean methylation levels were detected by a two-sample SAM statistic as implemented in the R package {\tt samr}. After Benjamini \& Hochberg correction for multiple testing, $2,770$ probes were found to be differentially methylated (DM) at the $5\%$ significance level. Of these, $620$ were also found in the differential subnetworks,  $90\%$ of which are concentrated in communities  $\mathcal{C}_2,\mathcal{C}_3,\mathcal{C}_5$ and $\mathcal{C}_6$. For example in community $\mathcal{C}_3$, there are 109 probes in total, half of which (54) are differentially methylated. Figure \ref{fig:diffmethy} shows the distribution of DM probes in the subnetworks. These results suggest that a differential analysis based exclusively on detecting mean levels of differential methylation may miss important differences that can only be identified by comparing the interaction networks.   

Table 2 provides a breakdown of the number of probes, differentially methylated probes ($q_i$), density ratio between control and case subnetworks ($R_i$), and distribution of enriched GO terms and KEGG pathways in the 6 communities (see also Figure \ref{fig:subnetworks}). Replicated GO terms and pathways involved in different communities were excluded in the subtotal. In $\mathcal{C}_5$ we found that all top 6 ranked significant BP terms were  related to interleukin-3 (IL-3), a  cytokine that is made by leukocytes and other cells in the body. IL-3 can increase the number of leukocytes, neutrophils, and platelets made by the bone marrow \cite{Diamantis89}. As Myelosuppression induced by chemotherapy is closely related to the effect of IL-3 in blood cells when suppressing a tumor during the therapy \cite{Dercksen93}, this may offer a possible explanation for the observed enrichment results. A possible explanation for the observed difference in the $\mathcal{C}_6$ cluster may be related to hypermethylation being linked to cancer \cite{Jones02, Iorio07}.
 
\begin{table}[h]
 \caption{\textbf{DNA co-methylation networks: a summary for different communities } }\label{community}
\begin{center}\begin{tabular}{lrrrrrrrr}\hline
 & $\mathcal{C}_1$ & $\mathcal{C}_2$ & $\mathcal{C}_3$ & $\mathcal{C}_4$ & $\mathcal{C}_5$ & $\mathcal{C}_6$ & subtotal & overall \\\hline
\# of probes & 418 &  66 &  109 & 34  & 347 &  200 & 1174 & 1642 \\ 
$q_i$ & 4 & 66&  54 & 1  & 338 &  97& 560 & 620 \\ 
$R_i$ & .181& .013& .012& 0& .002& 23.4& .145 &.156 \\
BP    & 320  & 25  &  38 &22  & 236  &  54 & 568  & 762 \\ 
MF  &54 & 4 & 15 & 3  & 43 &  27 & 125 & 154 \\ 
KEGG & 5 &  0 &  1 & 1  & 0 &  1 & 8 & 12 \\
 \hline
\end{tabular}\end{center}

\end{table}

\section{Conclusions} \label{discuss}

The comparison of DNA methylation or gene expression profiles across conditions is enabling the discovery of novel biomarkers for diagnosis or prognosis, and holds the promise to identify novel targets for therapeutical intervention. In this paper we have discussed the problem of comparing two labelled networks that are representative of two conditions (e.g. healthy and diseased tissues) and detecting statistically significant differences in their topology. Identifying disrupted interaction patterns in two labelled network comparisons is a challenging problem requiring novel statistical tools, and three contributions have been made here in this direction.  Firstly, we have proposed the GHD, a distance between two labelled networks that detects more subtle differences compared to the traditional Hamming distance. Secondly, we have demonstrated that the GHD can be used as a non-parametric test to assess whether two paired networks are statistically independent, and have described how p-values can be computed in closed-form without requiring computationally expensive permutation procedures. The plausibility of the conditions underpinning our derivations has been discussed using scale-free random network models as an example. Thirdly, we have proposed a fast subnetwork detection procedure, the dGHD algorithm, to detect localized topological differences between two paired networks.  This methodology provides a useful addition to standard two-sample tests that are commonly used for biomarker discovery. An initial evaluation has been carried out by comparing co-methylation networks inferred from healthy and cancer patients, and detecting differential subnetworks. Further experimental evaluation on independent datasets will be required to validate these results. In future work, the methodology could be extended to the case of more than two conditions.

\bibliographystyle{plain}
\bibliography{drrefs}      
\end{document}